\documentclass[prd, twocolumn, superscriptaddress, nofootinbib]{revtex4}
\usepackage{graphicx, epsfig}

\def\fun#1#2{\lower3.6pt\vbox{\baselineskip0pt\lineskip.9pt
  \ialign{$\mathsurround=0pt#1\hfil##\hfil$\crcr#2\crcr\sim\crcr}}}

\begin{document}

\title{Parameterization of Dark-Energy Properties: a Principal-Component
Approach}

\author{Dragan Huterer and Glenn Starkman}
\affiliation{Department of Physics, Case Western Reserve University, 
Cleveland, OH~~44106}

\begin{abstract}
Considerable work has been devoted to the question of how best to
parameterize the properties of dark energy, in particular its equation
of state $w$. We argue that, in the absence of a compelling model for
dark energy, the parameterizations of functions about which we have no
prior knowledge, such as $w(z)$, should be determined by the {\it
data} rather than by our ingrained beliefs or familiar series
expansions. We find the complete basis of orthonormal eigenfunctions
in which the principal components (weights of $w(z)$) that are
determined most accurately are separated from those determined most
poorly. Furthermore, we show that keeping a few of the
best-measured modes can be an effective way of obtaining information
about $w(z)$.
\end{abstract}

\maketitle

The discovery of the accelerating
universe~\cite{perlmutter-1999,riess} has been followed by a lot of
work concentrated on how best to parameterize dark energy and measure
its properties (\cite{huttur,weller,maor,PPF,SS} and references
therein). The parameters of choice have been the energy density in
this component, $\Omega_X$, and its equation-of-state ratio
$w(z)$~\cite{turner_white}. While current data allow interesting
constraints only if $w$ is assumed constant, future surveys, such as
those envisioned for the Supernova Acceleration Probe
(SNAP)~\cite{SNAP} or the Large-aperture Synoptic Survey Telescope
(LSST)~\cite{LSST} may allow relaxing that assumption and measuring,
for instance, the first derivative of $w$ with redshift, $w' \equiv
dw/dz|_{z=0}$. Despite these exciting prospects, it has become clear
that accurate constraints on a general $w(z)$ will remain elusive even
for the most ambitious surveys~\cite{huttur, weller}; hence, a simple
but general parameterization of $w(z)$ seems necessary.

Yet some issues remain obscure; in particular the question of what is
actually being measured by a given survey. In other words, what are
the quantitative weights on measurements of $w(z)$? Consider, for
example, a fiducial dark-energy model with the equation of state
$w(z)=-1+\sin(\pi z)$. Measurements of $w'$ would indicate a positive
quantity even though the redshift-averaged derivative of $w$ is zero;
this is because the largest sensitivity of a measurement is at the
low-redshift end where $w'(z)>0$. Using somewhat more general
arguments, it has been argued~\cite{huttur,weller} that $w(z)$ is
best determined around $z\sim 0.3$, and less accurately determined at
much lower and much higher redshift because of the Hubble law and
decreasing importance of dark energy respectively. We would like to
better understand this and, moreover, find a natural basis of weights
that represents the measurement.

We consider the coordinate distance $r(z)$ as the primary observable
of any survey (although this could easily be generalized to
number-count measurements, for example) and assume a flat universe
with energy density in matter $\Omega_M=0.3$ and that in the dark
component $\Omega_X=1-\Omega_M$.  We divide the redshift range of the
survey into $N$ bins centered at redshifts $z_i$ with corresponding
widths $\Delta z_i$ ($i=1, \ldots, N$).  We assume dark energy is
parameterized in terms of $w(z)$, which we define to be constant in
each redshift bin, with a value $w_i$ in bin $i$. In the limit
$N\rightarrow \infty$, this allows for a completely general $w(z)$. We
use $N=50$, which provides sufficient resolution yet doesn't require
large computational time (we have explicitly checked that the results
change little for $N\geq 20$). For piecewise-constant $w(z)$, the
energy density of the dark component evolves as (for $z$ in bin $j$,
$z_j-\Delta z_j/2 < z < z_j+\Delta z_j/2$)
\begin{eqnarray}
\rho_X (z)&= &\rho_X(z=0) 
\left (\frac{1+z}{1+z_{j}-\Delta z_{j}/2} \right
)^{3(1+w_j)} \times \nonumber\\[0.1cm]
&& \prod_{i=1}^{j-1} \left (\frac{1+z_i+\Delta
z_i/2}{1+z_i-\Delta z_i/2} \right )^{3(1+w_i)}.
\end{eqnarray}

Then $H^2(z)=H_0^2\,[(1-\Omega_M)\rho_X(z)/\rho_X(0)+\Omega_M(1+z)^3]$
and $H_0 r(z)=\int H_0/H(z')dz'$. Note that one could use the same
methodology to reconstruct other cosmological functions, for example
$f(z)\equiv H_0/H(z)$. 

In this paper, we only consider the function $w(z)$. Although $f(z)$
is easier to measure due to the fact that it is more directly related
to the observable luminosity distance, in order to understand the
behavior of dark energy, one needs to measure its evolution with time
and the scale factor --- therefore, measure a derivative of $\rho_X(z)$
or $f(z)$.  Since $w(z)$ is related algebraically to $df/dz$
\cite{reconstr}, one effectively needs $w(z)$. Therefore,
despite the fact that measurements of $f(z)$ or $\rho_X(z)$ will be
more accurate than those of $w(z)$, constraints on the latter quantity
will be crucial for understanding the nature of dark energy.

We use the Fisher matrix formalism to compute the covariance matrix
for the parameters $w_i$. We also include the ``nuisance parameter''
$\mathcal{M}$, which is an overall offset in the magnitude-redshift
diagram~\cite{perlmutter-1999} and $\Omega_M$; we then
marginalize over these parameters. For the fiducial survey, we assume
3000 type Ia supernovae (SNe) uniformly distributed in redshift
between $z=0$ and $z=1.7$. Any actual survey distribution will
undoubtedly be somewhat non-uniform in redshift, but we ignore this in
order avoid seeing the effects due to the distribution itself.

\begin{figure}[!t]
\includegraphics[height=3.5in, width= 2.4in, angle=-90]{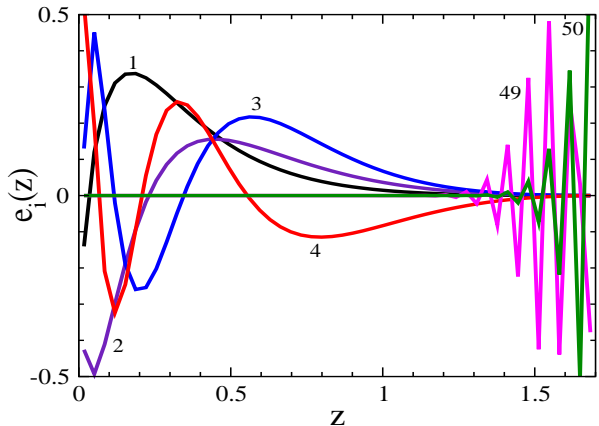}
\caption{The principal components of $w(z)$ for $\Omega_M$ perfectly known.
The four best-determined and two worst-determined eigenvectors are
shown and labeled for clarity. We have marginalized over the magnitude
offset $\mathcal{M}$. }
\label{fig:PC}
\end{figure}

\bigskip
{\it Principal components of the Fisher matrix.\hspace{0.2cm}} We start
by computing the Fisher matrix $F$ for our parameters $w_i$ ($i=1,
\ldots, N$), having marginalized over 
$\mathcal{M}$ and assuming for a moment that $\Omega_M$ has been
determined independently. We find the familiar result that the
accuracy in measurements of $w_i$, $\sqrt{(F^{-1})_{ii}}$, rapidly
increases with increasing $i$ (that is, redshift).  Now, it is a
simple matter to find a basis in which the parameters are
uncorrelated; this is achieved by simply diagonalizing the inverse
covariance matrix (which is in practice computed directly and here
approximated with $F$). Therefore
\begin{equation}
F  = W^T \Lambda W
\end{equation}

\noindent where the matrix $\Lambda$ is diagonal and rows of 
the decorrelation matrix $W$ are the eigenvectors $e_i(z)$, which
define a basis in which our parameters are
uncorrelated~\cite{hamilton}. The original function can be
expressed as 
\begin{equation}
w(z) = \sum_{i=1}^N \alpha_i\, e_i(z)
\label{eq:w_expand}
\end{equation}

\noindent  where $e_i$ are the 
``principal components''. Since $F$ is real and symmetric, $W$ can be
chosen to be orthogonal and of unit determinant, so that $\left\{
e_i(z) \right\}$ form an orthonormal basis. Using the orthonormality
condition, the coefficients $\alpha_i$ can be computed as
\begin{equation}
\alpha_i = \sum_{a=1}^N w(z_a)\, e_i(z_a).
\label{eq:coeff}
\end{equation}

Diagonal elements of $\Lambda$, $\lambda_i$, are the eigenvalues which
determine how well the parameters (in the new basis) can be measured;
$\sigma(\alpha_i)=\lambda_i^{-1/2}$. We have ordered the $\alpha$'s so
that $\sigma(\alpha_1)\leq\sigma(\alpha_2)\leq\ldots\leq\sigma(\alpha_N)$.

The principal components of $F$ are shown in Fig. \ref{fig:PC}. For
clarity, we only show 4 of the best-determined and 2 of the
worst-determined eigenvectors for $w(z)$ for the fiducial model
$w(z)=-1$.
The best-determined component of $w$ peaks at $z\approx 0.2$, enters with
the coefficient $\alpha_1=-4.18$, and can be measured to an accuracy
of 2.5\%.  We find that the eigenvectors depend weakly on the fiducial
cosmological model, while they depend somewhat more strongly on the
set of cosmological parameters; for example, marginalizing further
over $\Omega_M$ with a prior of 0.03, the best-determined eigenvector
of $w$ peaks at $z\approx 0.1$ while the others are mostly
unaffected. In addition, allowing that $\Omega_M$ is not precisely
known increases the uncertainty of how well the first few modes can be
measured by 50-100\%, while not affecting the higher modes
much. Finally, note that, although the best-measured modes peak at
relatively low redshifts, one still needs the full redshift range
($z\lesssim 2$) in order to measure those modes accurately.

Note a few nice things about this decomposition (which has recently
also been applied to the weak lensing reconstruction of the radial density
field~\cite{hu_keeton}). First of all, the measured eigenvectors are
orthonormal and ordered by how accurately they can be measured.  The
$M^{\rm th}$ best-determined eigenvector has precisely $M-1$ nodes,
which makes the interpretation quite natural: the first eigenvector
corresponds to the ``average of $w(z)$'', the second one to the
``first derivative of $w$'', the third one to the ``second derivative
of $w$'', etc. We argue that, in the absence of a compelling
theoretical argument for any particular parameterization of $w(z)$
except perhaps $w(z)\equiv-1$, these modes are the natural basis in
which the function $w(z)$ should be considered for any given
experiment.  Furthermore, the spread in accuracies is larger than
before: the best-determined modes are better determined than the
best-determined (original) components $w_i$, while the
worst-determined modes are determined more poorly than the
worst-determined $w_i$. Of course, the total constraint upon $w(z)$ at
any given redshift $z_a$
\begin{equation}
\sigma(w(z_a)) = \left (\sum_{i=1}^N 
\sigma^2(\alpha_i)\, e_i^2(z_a)\right )^{1/2}
\label{sig_w}
\end{equation}

\noindent does not depend on the chosen basis. Consequently, 
by exploiting only a few of the best-determined modes we might hope to
gain in accuracy while not biasing the result too much; we discuss
this later on.

\bigskip
{\it A ``sweet spot''?\hspace{0.2cm}} The best-determined eigenvector of
$w(z)$ peaks at $z\simeq 0.2$, although other parameter choices (not
including the magnitude offset $\mathcal{M}$, for example, or
marginalizing over $\Omega_M$) can change this to other values, even
$z\rightarrow 0$. At first this may seem at odds with the result that
$w(z)$ is typically best determined around $z\sim
0.3$~\cite{huttur,weller}. However, the question of where the ``sweet
spot'' -- the point of minimal uncertainty -- of $w(z)$ falls depends
on the function which is used to parameterize $w(z)$. For example,
using the parameterization $w(z)=w_0 + w' z$ leads to a sweet spot
around $z\sim 0.3$. Fitting a low-order polynomial to the distance
data may lead to one or two sweet spots in the fitted
function~\cite{huttur,weller}.  However, the equation-of-state ratio
cannot truly be isolated at a given redshift; moreover, there is no
compelling theoretical motivation for any particular
parameterization. We therefore argue that the weights we compute are a
better representation of the sensitivity of $w(z)$ than the sweet spot
in any particular parameterization.

\bigskip
{\it Testing the Constancy of $w(z)$.\hspace{0.2cm} } The only functional
form for $w(z)$ for which, we would argue, there is a strong
theoretical bias is a constant $w$, in particular $w(z)=-1$.  It is
therefore important to test whether the equation of state ratio $w$ is
constant or not. Non-constant $w(z)$ would point toward a dynamical
mechanism for late-time acceleration of the universe.
To test constancy of $w$ one could use actual
$r(z)$ data to compare $\chi^2$ for models of constant $w(z)$
vs. those of varying $w(z)$.  Or one could simply measure
$w'$~\cite{coorayhuterer, huttur, weller}.  Here we seek a
more general approach.

Although the $M$th eigenvector $e_M(z)$ roughly corresponds to the
$(M-1)$th derivative of $w(z)$, it unfortunately cannot be used
directly to test the constancy of $w$ simply because the coefficient
$\alpha_M$ is not zero even for $w(z)={\rm const}$.  Now, from
Eq.~(\ref{eq:coeff}) it follows that, for constant $w(z)\equiv w_c$
\begin{equation}
\bar{\alpha_i} = w_c\,\sum_{a=1}^N e_i(z_a).
\end{equation}

\noindent where $\{\bar\alpha_i\}$ correspond to the hypothetical $w_c$.
One can then perform a simple $\chi^2$ test to determine whether
$w$ is constant:
\begin{equation}
\chi^2=\sum_{i=1}^M {(\alpha_i-\bar{\alpha_i})^2
\over \sigma^2(\alpha_i)},
\end{equation}

\noindent where we have chosen to keep only the first $M$ eigenmodes, since the
best-determined modes will contribute the most to the sum (the test is
valid regardless of the value of $M$). Since the best-fitting $w_c$ is
to be determined by finding the minimum $\chi^2$, there are $M-1$
degrees of freedom. For example, for a fiducial model
$w(z)=-1.0+0.3\,z$, keeping $M=4$ modes rules out constant $w$ at
98.7\% CL. For comparison, the standard $w_0$-$w'$ test applied to
this fiducial model rules out constant $w$ at about $98.5\%$ CL, since
$\sigma(w')\approx 0.13$. Although the two tests give comparable
results, the former describes $w$ with four parameters rather than two
and is therefore more general than the $w'$ test.  In particular, the
proposed test has a nice feature that the number of principal
components that are used, $M$, can be chosen so as to maximize the
strength of the desired constraint.

\bigskip
{\it Function Reconstruction?\hspace{0.2cm}} From Fig.~\ref{fig:PC} it
is apparent that the well-determined eigenmodes: (1) are non-zero at
low end of the redshift range and zero at the high end, and (2) do not
oscillate much. The opposite is true for the poorly determined
eigenvectors. This tells us that the accuracy in determining $w(z)$ is
best if this function is smooth {\it and} if we are trying to
determine it at low to moderate redshift ($z\lesssim 1$).

We have seen above that the eigenmodes of $w(z)$ are roughly
ordered by the absolute size of their coefficients -- more noisy modes
contribute less to $w(z)$. One can then use the first $M$ eigenvectors
(those measured most accurately) in order to
approximately reconstruct $w(z)$:
\begin{equation}
w(z) = \sum_{i=1}^N \alpha_i\, e_i(z)
\approx 
\sum_{i=1}^M \alpha_i\, e_i(z)
\end{equation}

\noindent where $M\leq N$. 
Obviously, in the case of $M=N$ we recover the original error bars,
which are typically hopelessly large (at least for $N\gtrsim{\rm
few}$, which gives sufficiently high resolution in redshift). When
$M<N$, two things happen: $w(z)$ is reconstructed less accurately (so
the bias increases), but the error bars are smaller (so the variance
decreases). Indeed, getting rid of the noisy eigenmodes corresponds to
setting $w(z)$ at high redshift end to zero, which will bias any
$w(z)$ that doesn't actually go to zero at that redshift.  

\begin{figure}[!t]
\includegraphics[height=1.8in, width=1.3in, angle=-90]{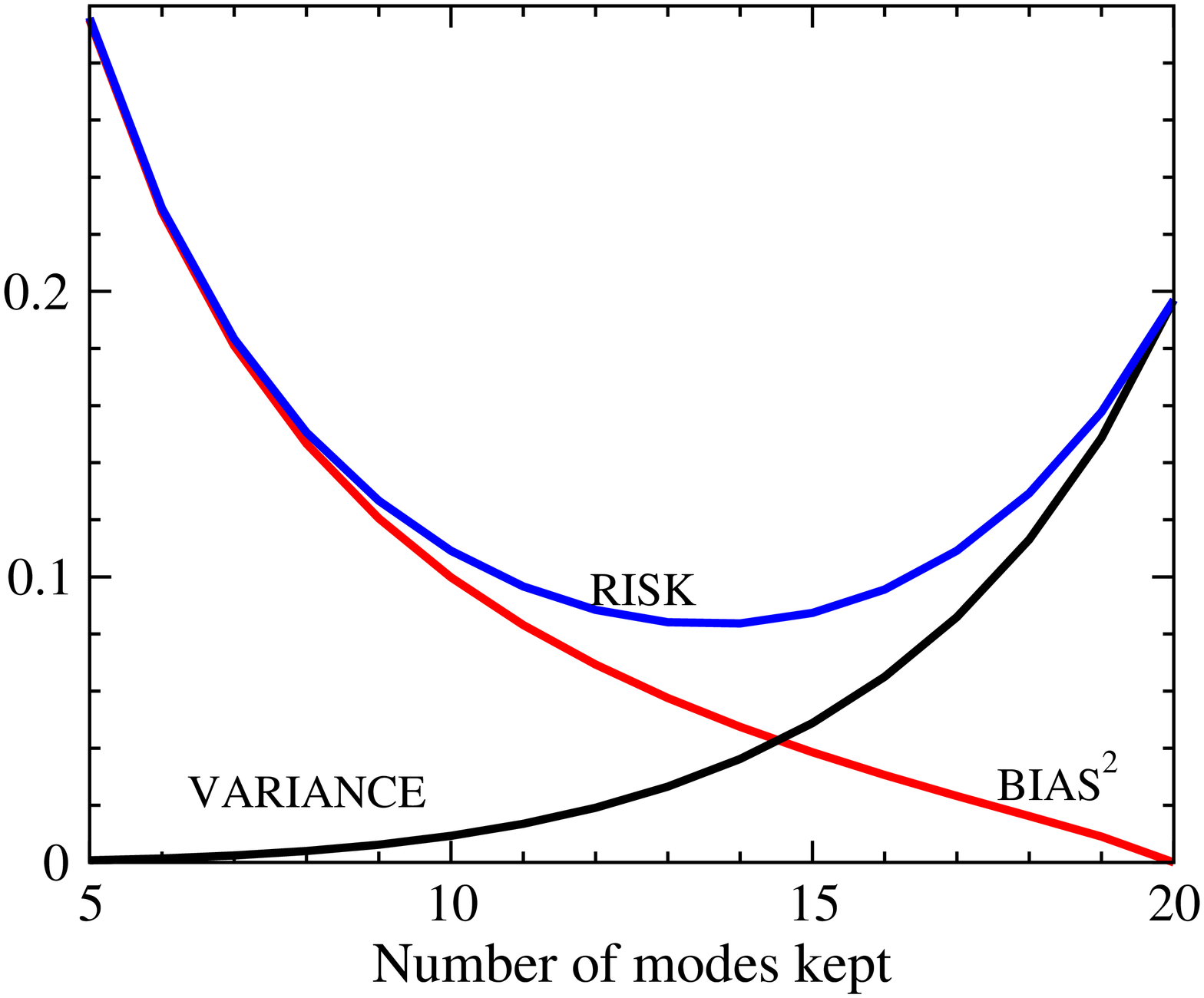}
\\[-0.15cm]
\includegraphics[height=1.8in, width=1.5in, angle=-90]{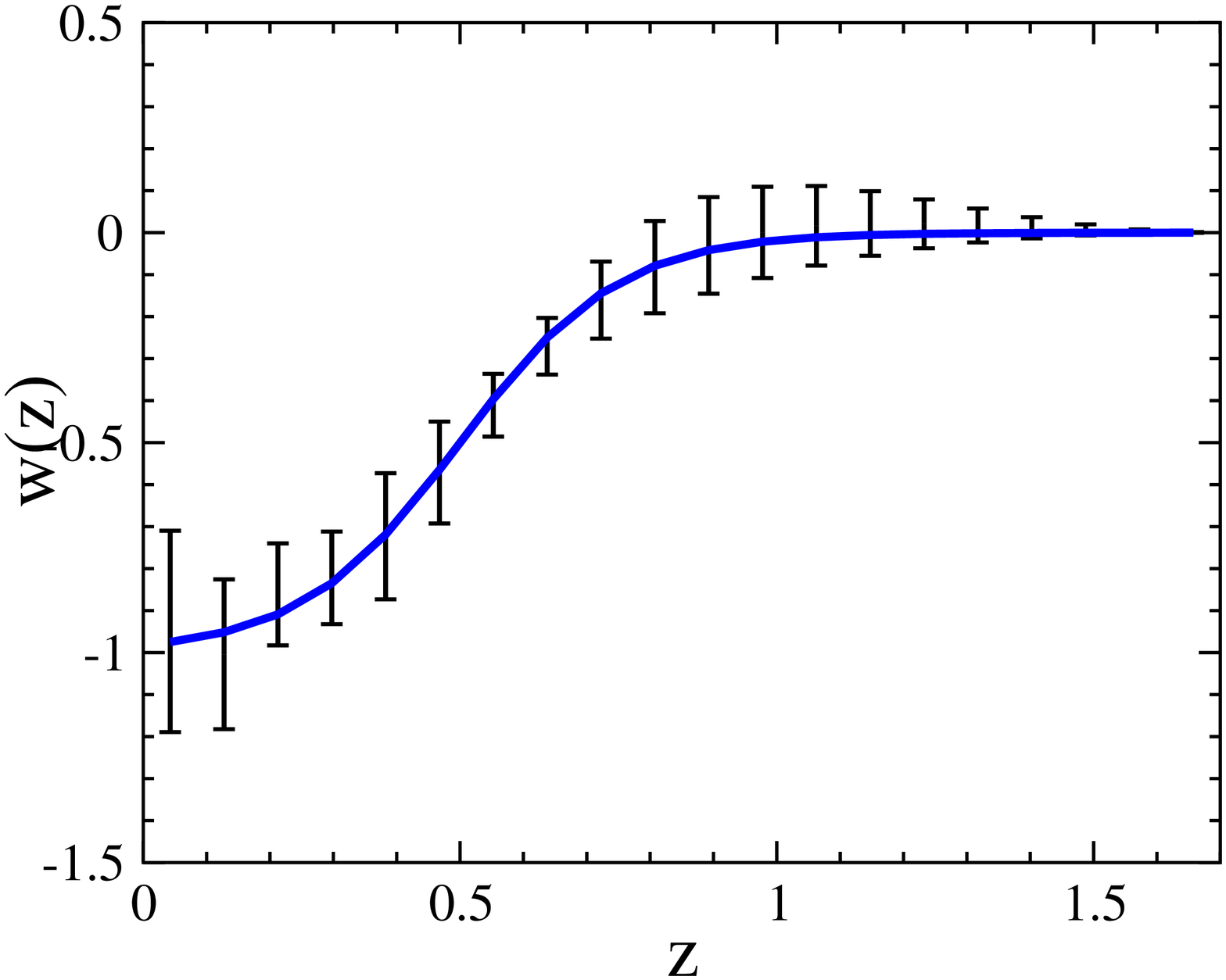}\nobreak
	\hspace{-0.4cm}
\includegraphics[height=1.8in, width=1.5in, angle=-90]{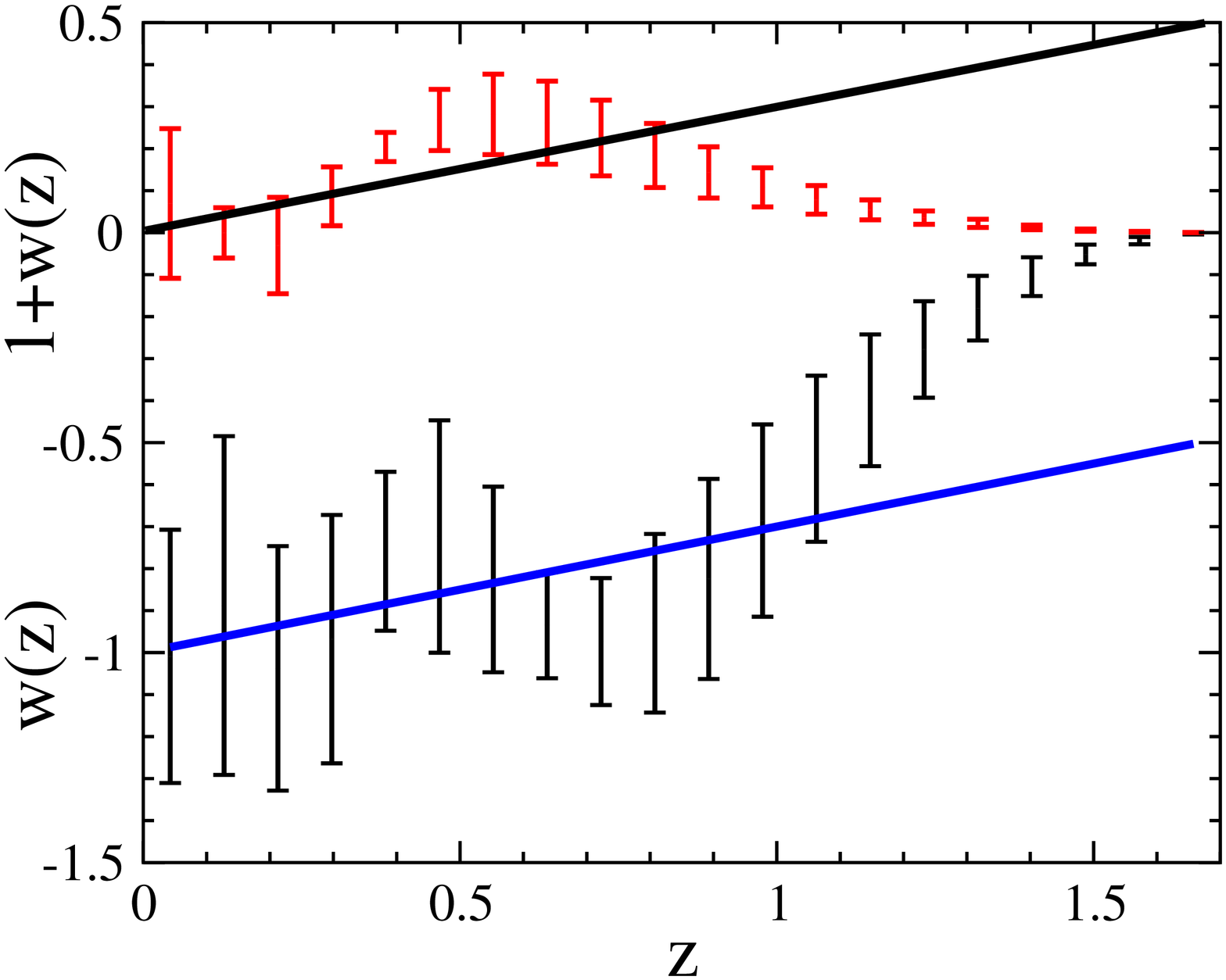}
\caption{Reconstruction of $w(z)$ by keeping only a fraction of
eigenvectors so as to minimize risk. Top panel: illustration of the
minimization of risk. Bottom left: optimal reconstruction (68\%
C.L. error bars shown) of fiducial $w(z)$ (solid line) that goes to
zero at high-redshift end.  Bottom right: optimal reconstruction of
$w(z)$ that does not go to zero at high-redshift end. Also shown
on this panel is optimal reconstruction of $1+w(z)$ for the same $w(z)$ model. 
}
\label{fig:reconstr}
\end{figure}

Fig.~\ref{fig:reconstr} illustrates these arguments (we choose $N=20$
for definitiveness).  To choose the optimal number of eigenmodes to be
kept, $M$, the statistically correct thing to do is to {\it minimize
risk}~\cite{wasserman}, where
\begin{eqnarray}
{\rm risk} &=& {\rm bias}^2 + {\rm variance} \\
&=& \sum_{i=1}^N \left (w(z_i)-\bar{w}(z_i)\right )^2 +
\sum_{i=1}^N  \sigma^2(w(z_i)),
\end{eqnarray}

\noindent the sums in Eqs.~(\ref{eq:w_expand}) and (\ref{sig_w}) run from
1 to $M$, and $\bar{w}$ is the fiducial value of $w$. The top panel of
Fig.~\ref{fig:reconstr} illustrates the minimization of risk.  The
bottom left panel of the Figure shows that the fiducial
$w(z)=1/2(-1+\tanh(z-0.5))$, which smoothly asymptotes to zero at
$z\gtrsim 0.8$, can be reconstructed without much bias; its risk is
minimized by keeping $M=4$ terms. The same is not true for
$w(z)=-1+0.3\,z$ (bottom right; $M=5$ for minimum bias), which doesn't
go to zero within our redshift range and is therefore reconstructed
with a large bias. Furthermore, the reconstructed quantities that are
linearly related to $w(z)$, such as $1+w(z)$, will also go to zero at
high redshift if $M<N$ (bottom right panel).  The resulting bias in
$w(z)$ is therefore different for the two cases, and depends on
whether $w(z)$ or, say, $1+w(z)$ is reconstructed.

This analysis shows that, while reconstructing $w(z)$ at $z\lesssim 1$
is somewhat promising, it does not seem possible to recover $w(z)$ at
$z>1$ in a truly model-independent way.  We have checked that even
adding a large number of SNe at the high-redshift improves the
reconstruction only marginally, leaving a significant bias at the high
redshift end. The fact that bias (and therefore risk) will be
difficult to estimate in practice exacerbates the problem. Note too
that the proposed reconstructions of the equation of state
ratio~\cite{reconstr, huttur, weller} are parametric, since they use a
fitting function to fit the distance-redshift data.  While planned
surveys may provide accurate determinations of the fitting parameters,
they cannot test well the validity of the survey parameterization.

\bigskip
{\it Conclusions.\hspace{0.2cm}} A number of models that roughly
explain the observable consequences of dark energy were proposed in
recent years, some of them starting from fundamental physics and
others being purely phenomenological. However, the origin and nature
of dark energy remain unknown, and it is safe to say that none of the
models should be taken too seriously at this point. Therefore, it
seems wise to approach the question of constraining the properties of
dark energy empirically, with as few prior assumptions as
possible.

Here we propose that rather than using various parameterizations
proposed in the literature to describe dark energy (which typically
parameterize the equation of state $w(z)$ using series expansions
etc.), one can simply let data decide which weights of these functions
are measured best, and which ones are measured most poorly.  These
weights are the natural basis that parameterizes the measurements of
any particular survey.

For definitiveness, we assumed a cosmological distance-redshift survey
containing 3000 SNe uniformly distributed in redshift.  We computed
the weights of $w(z)$ and showed that accurately measured modes
(weights) are rather smooth and go to zero at higher redshifts, while
the opposite is true for poorly measured modes.  The
previously-considered ``sweet spot'' in the sensitivity of $w(z)$ is
largely a function of the choice of parameterization, and the shape of
the first principal component is a better indicator of the redshift(s)
at which $w$ is being measured.

With the proposed parameterization, the test of whether $w(z)$ is
constant is straightforward and intuitive. Although the reconstruction
of $w(z)$ is straightforward to implement, the reconstructed $w(z)$ is
noisy, and it is advantageous to keep only the best-measured modes in
order to decrease the statistical reconstruction error. This
introduces a systematic bias at $z\gtrsim 1$ for most models, roughly
independently of the redshift coverage of SNe.  We therefore conclude
that while model-independent statements about $w(z)$ at $z\lesssim 1$
may be feasible, those at $z\gtrsim 1$ will be unreliable.

In our opinion, the greatest advantage of this approach is simply
having an intuitive and quantitative answer as to what is actually
being measured by a given survey. The next logical step is to apply
this method to other cosmological tests.  We will address this in a
future publication.

\vspace{-0.3cm}
\begin{acknowledgments}
\vspace{-0.35cm}
We thank Wayne Hu, Eric Linder and Saul Perlmutter for useful
conversations, and Max Tegmark for insightful comments on the
manuscript.  This work is supported by a Department of Energy grant to
the particle astrophysics theory group at CWRU.
\end{acknowledgments}

\end{document}